\documentclass[a4paper,11pt]{article}
\pdfoutput=1
\usepackage{jcappub}
\usepackage[T1]{fontenc}
\usepackage{subfig}
\usepackage{amsmath}
\usepackage[normalem]{ulem}
\usepackage{graphicx}% Include figure files
\usepackage{bm}% bold math
\usepackage{natbib}
\usepackage{color}
\usepackage[utf8]{inputenc}

\def\ltsima{$\; \buildrel < \over \sim \;$}
\def\simlt{\lower.5ex\hbox{\ltsima}}
\def\gtsima{$\; \buildrel > \over \sim \;$}
\def\simgt{\lower.5ex\hbox{\gtsima}}
\graphicspath{{plots/}} 
 
\def\[{\begin{equation}}
\def\]{\end{equation}}

\author[1,2]{Nicola Bellomo,}
\emailAdd{nicola.bellomo@icc.ub.edu}
\author[1,3]{Emilio Bellini,}
\emailAdd{emilio.bellini@physics.ox.ac.uk}
\author[1,4]{Bin Hu,}
\emailAdd{binhu@icc.ub.edu}
\author[1,5]{Raul Jimenez,}
\emailAdd{raul.jimenez@icc.ub.edu}
\author[6,7]{Carlos Pena-Garay,}
\emailAdd{penya@ific.uv.es}
\author[1,5]{Licia Verde}
\emailAdd{liciaverde@icc.ub.edu}

\affiliation[1]{ICC, University of Barcelona (UB-IEEC), Marti i Franques 1, 08028, Barcelona, Spain.}
\affiliation[2]{Dept. de Fisica Quantica i Astrofisica, Universitat de Barcelona, Marti i Franques 1, E08028 Barcelona, Spain}
\affiliation[3]{University of Oxford, Denys Wilkinson Building, Keble Road, Oxford, OX1 3RH,  UK}
\affiliation[4]{Department of Astronomy, Beijing Normal University, Beijing, 100875, China}
\affiliation[5]{ICREA, Pg. Lluis Companys 23, 08010 Barcelona, Spain.} 
\affiliation[6]{Instituto de Fisica Corpuscular, CSIC-UVEG, P.O.  22085, Valencia, 46071, Spain.}
\affiliation[7]{Laboratorio Subterr\'aneo de Canfranc, Estaci\'on de Canfranc, 22880, Spain.}

\date{\today}

\title{Hiding neutrino mass in modified gravity cosmologies}

\abstract{Cosmological observables show a dependence with the neutrino mass, which is partially degenerate with parameters of extended models of gravity. We study and explore this degeneracy in Horndeski generalized scalar-tensor theories of gravity. Using forecasted cosmic microwave background and galaxy power spectrum datasets, we find that a single parameter in the linear regime of the effective theory dominates the correlation with the total neutrino mass.  
For any given mass, a particular value of this parameter approximately cancels the power suppression due to the neutrino mass at a given redshift. The extent of the cancellation of this degeneracy depends on the cosmological large-scale structure data used at different redshifts. We constrain the parameters and functions of the effective gravity theory and determine the influence of gravity on the determination of the neutrino mass from present and future surveys.}

\begin{document}
\maketitle
\flushbottom

\section{Introduction}

Recent cosmological observations have brought upon us the era of precision cosmology. To challenge the current standard cosmological model seems to require very precise cosmological parameters determinations, as all current available observations are consistent with the simplest $\Lambda$CDM model \cite{Ade:2015xua}. However, the current model is just a convenient phenomenological description of the Universe as it gives no insight on the nature of the individual energy components like dark matter and dark energy. Most likely before major breakthroughs in our understanding of the Universe come through, precision cosmology should verify yet undetected, small effects, corresponding to standard expectations. Among them, the effect of neutrino masses on large-scale structure is the most promising candidate to verify cosmology at the sub-percent level. Is there any chance for surprise? This has been addressed in a series of works which involve a plethora of modified cosmological models (for a review, see \cite{Ade:2015rim,Clifton:2011jh}) where some specific piece of the extended model mimics the impact of neutrino masses. Modified gravity models stand as the most promising alternative to the current paradigm (see e.g., \cite{Kristiansen2010609,lavacca:cdmdeneutrinos,lavacca:ddeneutrinos}).

The impact of deviations of Einstein gravity on the determination of neutrino masses has been studied and analysed both in the linear and nonlinear regime \cite{Baldi:2013iza,Barreira:2014ija,Shim:2014uta}. Most often, the extended gravity models are specific and simplified scenarios which avoid the exploration of large parameter spaces in time-consuming simulations and/or analysis. The outcome of these studies typically shows a qualitative understanding of the influence of the modified model's parameters in the adopted neutrino mass bound. 

In this paper, we make a more general characterization of the influence of modified gravity models on the determination of the neutrino mass. We characterize and analyse fully general massive neutrino scalar-tensor (Horndeski) cosmologies, for the first time, working with the effective theory and observations in the linear regime \cite{Horndeski:1974wa}. The modified gravity models are very generally characterized by a minimal number of given functions, set by a limited number of parameters. 
The redshift dependence of these functions is driven by searching for the largest impact on the neutrino mass constraints. In this framework, we can address the questions: ``where is the degeneracy between neutrino mass and a modified gravity model hidden?'' and ``how could it be partially resolved?''. Moreover, we can quantitatively characterize the knowledge of neutrino mass in the general models under scrutiny. Our results can be directly applied to theoretically motivated tensor-scalar gravity theories by matching the functions of the effective theory to those used here (see \S \ref{sec:models}).

\section{Models}
\label{sec:models}

The idea that DE/MG models could hide the effects of the mass of neutrinos on cosmological scales is intriguing and deserves investigation. However, one has to choose carefully the framework to work with. A simple DE/MG model has not enough freedom to be used both to drive the expansion history and to affect the formation of cosmic structure as massive neutrinos do. The reason is that neutrinos become non-relativistic at typical times ($z\simeq 100$) far before the usual on-set of DE created to drive the late-time cosmic acceleration ($z\simeq 1$). Then, the models we look at should have at least two different time scales, one for the background and one for the perturbations. Our focus is on a broad class of scalar-tensor theories, namely the Horndeski class of models \cite{Horndeski:1974wa,Deffayet:2009mn,Kobayashi:2011nu}. Horndeski is the most general theory with one extra scalar propagating degree of freedom that have second-order equations of motion on any background and that satisfies the weak equivalence principle, i.e.\ all matter species are coupled minimally and universally to the same metric $g_{\mu\nu}$. This class of models has the freedom to choose four arbitrary functions of two variables, i.e.\ the extra propagating degree of freedom $\phi$ and its canonical kinetic term $X=-\phi^{;\mu}\phi_{;\mu}/2$. Any choice of these free functions affects simultaneously both the expansion history and the evolution of the perturbations.

A different approach, still encoding all the freedom of the Horndeski class of models, is the so called Effective Field Theory (EFT) for Dark Energy \cite{Gubitosi:2012hu,Bloomfield:2012ff,Gleyzes:2013ooa,Hu:2013twa}. In \cite{bellini:horndeski} it was noticed that all the amount of cosmological information up to linear order in perturbation theory in Horndeski, can be compressed into one function of time driving the expansion history of the universe (the Hubble parameter $H(t)$), plus four functions of time and one constant acting just at the level of the perturbations. The constant can be identified as the fractional density of matter today ($\Omega_{m0}$) and the other functions of time have been dubbed: \textit{kineticity} $\alpha_K(t)$, \textit{braiding} $\alpha_B(t)$, \textit{Planck mass run-rate} $\alpha_M(t)$ or equivalently the \textit{Planck mass} $M_*(t)$, where $\alpha_M\equiv \frac{d\ln M_*^2}{d\ln a}$, and \textit{tensor speed excess} $\alpha_T(t)$. The advantage of using this approach instead of the original Horndeski function is twofold: (i) since all the cosmological information is compressed into a minimal set of functions, it is easier to understand the phenomenology of the models we are studying, and (ii) we can separate the contributions to the background from the contributions to the perturbations. In other words, we can directly modify the evolution of the perturbations keeping the expansion history fixed and compatible with data.

The price one has to pay for using this approach is that, since it is not possible with current data (and probably also with future ones) to constrain the $\alpha_i$ functions non-parametrically, any parametrization we choose can not be considered as representative of the full parameter space of the Horndeski theories, but it refers to specific and possibly fine-tuned class of models. In particular, it is not trivial to link this phenomenological description with classes of action-based theories \cite{Linder:2015rcz}. Nevertheless, our approach is still useful since our purpose is to give a proof of principle that, under particular circumstances, the effects on the observables of the mass of neutrinos can be hidden into the gravity sector.

As stated before, the class of models we should look at, must have two time scales, one related to the on-set of DE at the background level, and one that can mimic the transition of neutrinos from the relativistic regime to the non-relativistic one. Then, we fix the expansion history to be the one predicted by the standard $\Lambda$CDM in a flat universe. On the other hand, the $\alpha_i$ functions are parametrized as follow: $\alpha_K(z)=c_K$, $\alpha_B(z)=c_B\times\texttt{mod}(z)$, $\alpha_M(z)=c_M\times\texttt{mod}(z)$ and $\alpha_T(z)=c_T\times\texttt{mod}(z)$, where $c_j$ are constants. The choice of a constant $\alpha_K$ is due to the fact that its effect is subdominant on the growth of structures w.r.t.\ the other alphas and it is poorly constrained by present data \cite{bellini:constraints} but also the next generation of surveys will not bring its uncertainty down to the level of the other alphas \cite{Alonso:2016suf}. The function \texttt{mod}(z), is defined ad-hoc to switch on modifications to Einstein gravity at a given redshift $z_{th}$ with a transition given by $\Delta z$. Then, a convenient formulation can be
\begin{equation}
\texttt{mod}(z)=\frac{1+\tanh\left(\frac{z_{th}-z}{\Delta z}\right)}{1+\tanh\left(\frac{z_{th}}{\Delta z}\right)}.
\label{equation_modulation_function}
\end{equation}
For our purposes, we choose $z_{th}$ close to the redshift of neutrinos becoming non-relativistic and $\Delta z$ comparable to $z_{th}$ (see other scenarios in \cite{perenon:horndeski}).

In addition, Horndeski theories introduce a new scale dependence, the braiding scale $k_B$, that can be useful to separate the effects on small and on large scales. Indeed, $k_B$ signals the transition between two different gravity regimes, imprinting a characteristic scale-dependence in the Power Spectrum. In a $\Lambda$CDM  background, it reads
\begin{equation}
\frac{k_B^2}{a^2H^2}=\frac{9}{2}\Omega_m + 2\left(\frac{3}{2}+\frac{\alpha_K}{\alpha_B^2}\right)\left(\alpha_M-\alpha_T\right).
\label{equation:braiding_scale}
\end{equation}
While the braiding scale is related to the scale where the shape-dependent modification of the growth manifests itself, in the $\Lambda$CDM model this scale is undefined. For this reason we cannot report a  $k_B$ value or limit where one univocally recovers the standard gravity regime.

\section{Mock data and Likelihoods}
\label{sec:data}
We analyse our modified gravity models against the CMB temperature, E-mode polarization and deflection angle power spectra, as well as their cross correlation, simulated with the fiducial cosmological model using {\it Planck} results \cite{Planck:fiducial} and {\it Planck} blue book beam and noise specifications \cite{Planck:2006aa}. We use the cosmological parameters listed in Table \ref{tab:parameters} along with the CAMB code \cite{Lewis:1999bs} to produce the fiducial CMB temperature and E-mode polarization power spectra. We feed these to the FuturCMB package \cite{Perotto:2006rj} to compute the noise power spectra for T, E-modes and the lensing deflection angle based on the Hu-Okamoto~\cite{Hu:2001kj} quadratic estimator. For further details about the FuturCMB code we refer the reader to \cite{Perotto:2006rj} while for the construction of the spectrum likelihood we refer to \cite{%Tegmark:1996bz,
Lewis:2005tp,Perotto:2006rj}. 
\begin{table}
\centering
\begin{tabular}{|c|c|c|}
\hline
\bf{MCMC described in Sec.}	&	\bf{Fixed parameters}	&	\bf{Varying parameters}	\\
\hline
Hiding neutrino masses	&	$c_K,\Sigma m_\nu,z_{th},\Delta z$	&	$\omega_b,\omega_{cdm},h,A_s,n_s,z_{reio},c_B,c_M,c_T$	\\
Removing degeneracies	&	$c_K,c_T,z_{th},\Delta z$	&	$\omega_b,\omega_{cdm},h,A_s,n_s,z_{reio},c_B,c_M,\Sigma m_\nu$	\\
\hline
\end{tabular}
\caption{List of fixed and varying parameters in MCMC runs described in sections "Hiding neutrino mass" and "Removing degeneracy". We assumed a flat prior for each varying parameter.}
\label{tab:MCMC}
\end{table}
\begin{table} [htb]
\centering
\begin{tabular}{ |c|c|c|c| } 
\hline
 	& fiducial  & MCMC run	&	MCMC run	\\ 
\hline
\bf{Parameter}   	& \texttt{GR$_\mathrm{fid}$+M0}   &  \texttt{MG$_{1}$+M500}	&	\texttt{MG$_{2}$+M500}	\\ 
\hline
$100\omega_b$ & 2.218 & [2.23,2.28]	&	[2.26,2.30]	\\
$\omega_{cdm}$ & 0.1205 & [0.113,0.115]	&	[0.112,0.114]	\\
$h$ & 0.6693 & [0.650,0.659]	&	[0.656,0.665]	\\
$10^9A_s$ & 2.124 & [2.08,2.15]	&	[2.52,2.63]	\\
$n_s$ & 0.9619 & [0.969,0.979]	&	[0.965,0.975]	\\
$z_{reio}$ & 8.24 & [7.7,9.3]	&	[15.3,17.0]	\\
$c_K$	&&	10	&	1000	\\
$10^2c_B$ &&[3.7,5.7]	&	[3.7,5.7]	\\
$10^3c_M$ &&[3.1,7.9]	&	[2.8,7.4]	\\
$10^2c_T$ &&[-0.8,-0.4]	&	[-15.5,-10.1]	\\
$z_{th}$	&&	100	&	100	\\
$\Delta z$	&&	20	&	20	\\
\hline
\end{tabular}
\caption{Cosmological parameters of the fiducial model and the 95 \% CL intervals of the modified gravity model parameters for two particular cases that have different values of $c_K$. These results were derived by analysing \textit{CMB + Full P(k)}$(\sigma_\mathrm{obs})$ dataset. The apparent detection of a non-zero MG parameter is due to the fact that cosmological parameters alone can not compensate for the presence of significantly heavy neutrinos.}
\label{tab:parameters}
\end{table}

We also consider an Euclid-like survey \cite{euclid:databook}, with 14 redshift bin in the range $[0.7,2.0]$ and scales in the range $[0.001,0.12]\ \mathrm{Mpc^{-1}}$ (or $k<0.2h$/Mpc), in order to stay within the linear regime at every redshift. The number of galaxies per square degree is given by the Euclid 2016 settings, as well as the observed fraction of the sky. Our  basic likelihood was constructed following \cite{audren:forecast} and includes a scale independent bias, a Kaiser term for redshift space distortions and errors in determining galaxies line of sight positions, which contains spectroscopic/photometric errors and Fingers of God effects. This basic setting will be indicated as \textit{P(k)} dataset. We have then extended it to the \textit{Full P(k)} dataset, where we added information due to geometrical distortions (i.e. BAO and Alcock-Paczynski) related to different expansion histories. In both cases we can choose to include just an observational error $\sigma_{obs}$, given by shot noise and cosmic variance, or also a theoretical error $\sigma_{the}$, as explained in Appendix A of \cite{audren:forecast}, which should account for all possible effects not considered in our likelihood.
To calculate cosmological observables in our analysis we have used the \texttt{hi\_class} \cite{zumalacarregui:hiclass} public code, an extension of the \texttt{CLASS} \cite{lesgourgues:class,blas:class,lesgourgues:class_ncdm} code that allows us to include an additional scalar degree of freedom in the gravitational sector and to model its effects on gravity and matter, while the Monte Carlo Markov Chain (MCMC) forecast has been done with the \texttt{MONTEPYTHON} \cite{audren:montepython} code. Chains were considered converged when the Gelman and Rubin parameter was $R-1<0.01$.

In this work we used the following neutrino models: a) three massless neutrinos (model \texttt{M0}), b) three massive neutrinos in the normal ordering with $\sum m_\nu=150\ meV$ (model \texttt{M150}) and c) three quasi-degenerate neutrinos in the normal ordering with $\sum m_\nu=500\ meV$ (model \texttt{M500}). The first and second model were used to create the fiducial mock data, while the third was used along with modified gravity models.

\section{Hiding neutrino masses}
We have run several MCMC using the \textit{CMB+Full P(k)$(\sigma_\mathrm{obs})$} dataset (see Table \ref{tab:MCMC}). The fiducial model was calculated assuming General Relativity plus three massless neutrinos, using fiducial values reported in Table \ref{tab:parameters}. This fiducial will be indicated in what follows by \texttt{GR$_{\rm fid}$+M0}. Every MCMC run was characterised by a different choice of the values of the kineticity function $\alpha_K$ and of the modulation function parameters $(z_{th},\Delta z)$, which were taken as fixed parameters. We explored the parameter space given by the cosmological and MG parameters $\{\omega_b,\omega_{cdm},h,A_s,n_s,z_{reio},c_B,c_M,c_T\}$, assuming a flat prior for each of them. In every run the neutrino mass scheme was given by the \texttt{M500} model. The results for the particular case \texttt{MG$_{1}$+M500}, where $\alpha_K=10$, $z_\mathrm{th}=100$ and $\Delta z=20$, are shown in Table \ref{tab:parameters}.

Comparing the \texttt{MG$_{1}$+M500} model results with the fiducial model, we find a lower value of \textit{h}, required in order to preserve the acoustic peak scale, as well as a higher value of \textit{n$_s$}, as expected since it helps to compensate the neutrino power suppression at small scales. The value of $\omega_{cdm}$ decreases significantly in order to keep the global $\omega_m$ similar to the fiducial one, since also neutrinos contribute to the physical matter density with $\omega_\nu=0.0054$. We found compatibility between cosmological parameters confidence regions for different sets of the modulation function parameters. The main changes between the models due to the choice of $(z_{th},\Delta z)$ is the allowed range of MG parameters, in particular $\alpha_B$, since we have checked that it is responsible for the enhanced structure growth at small scales. Any deviation from GR that appears earlier requires a smaller global amplitude, since  the modifications of gravity will be at work for a longer period of time; on the other hand, modifications that become significant  slowly (or later) will require a bigger overall magnitude. The parameter $\alpha_M$ is highly correlated with $\alpha_B$ because they have opposite effects on the matter power spectrum. Notice that we have imposed stability conditions on the positiveness of the scalar field sound speed which partially induces such correlation.

\begin{figure}
\centerline{
\includegraphics[scale=0.6]{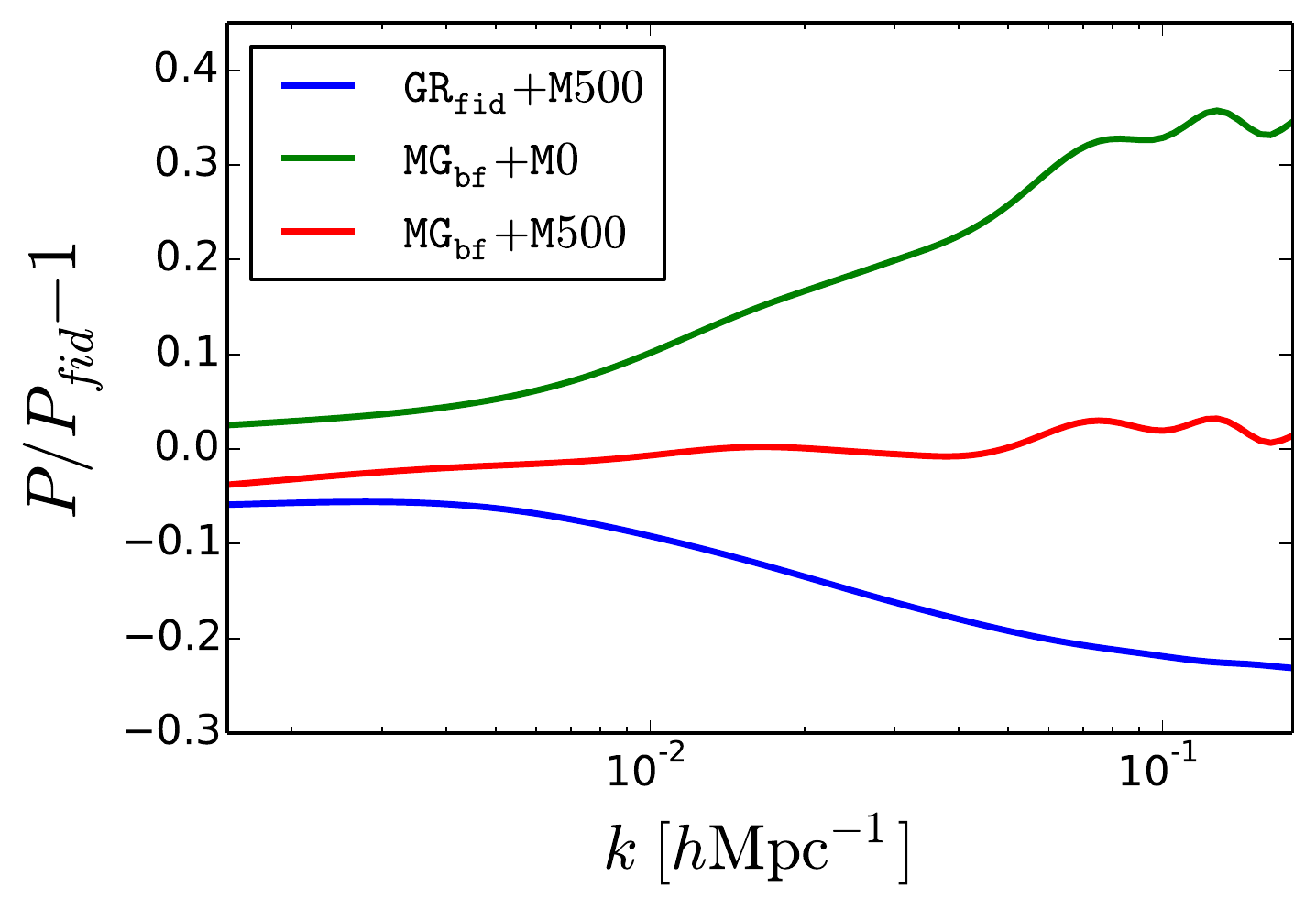}} % 80mm
\caption{Relative difference in the matter power spectrum of three models with respect to the fiducial model GR$_\mathrm{fid}$ at redshift $z=1.4$. The blue line shows the suppression due to  massive neutrinos \texttt{M500} model vs \texttt{M0}. The green line corresponds to the MCMC best fit cosmological and MG parameters with the \texttt{M0} neutrino model, showing the MG enhancement in the matter power spectrum at small scales. The red line represents the MCMC best fit of the MG model with massive neutrinos, which mimics the GR model with massless neutrinos.
\label{figure:step1_power_spectrum}}
\end{figure}

The potential of modified gravity to hide neutrino mass is shown in Figure \ref{figure:step1_power_spectrum}. Using the \texttt{MG$_{1}$+M500} result, we plot the relative matter power spectrum (to the GR$_\mathrm{fid}$) at representative redshift $z=1.4$ (at other redshift such as 0.5 or 2 the plot is virtually indistinguishable) for three different models: the fiducial model with massive neutrinos (\texttt{GR$_\mathrm{fid}$+M500}), the modified gravity best fit model with massless neutrinos (\texttt{MG$_\mathrm{bf}$+M0}) and the modified gravity best fit model with massive neutrinos (\texttt{MG$_\mathrm{bf}$+M500}). Massive neutrinos imprint a power suppression at small scales in the GR cosmology, while the modified gravity model with massless neutrinos leads to an increased power at small scales. Both effects, massive neutrinos and modified gravity, can be approximately cancelled as shown by the model \texttt{MG$_\mathrm{bf}$+M500}, where the relative differences are below the error bars. We stress here the generality of this result, which we have verified for several modulation function parameters and redshifts.

Contrary to naive expectations, $\alpha_K$ produces a detectable effect. In the \textit{CMB} only analysis, $\alpha_K$ changes the overall width of the MG parameters posteriors, but does not modify cosmological parameters. Once we include the \textit{Full P(k)}$(\sigma_{obs})$ dataset, cosmological parameters posteriors also change. Comparing the \texttt{MG$_{1}$+M500} to the \texttt{MG$_{2}$+M500} model results reported in Table \ref{tab:parameters}, we find that increasing values of $\alpha_K$ produce a substantial shift in the confidence interval of $A_s$ and, consequentially, of $z_{reio}$.

This shift can be attributed to the interplay between the two relevant scales, the neutrino free-streaming scale at the non-relativistic transition $k_{nr}$ and the MG braiding scale $k_B$. In order to hide the typical step-like feature in the power spectrum due to massive neutrinos, our MG model should introduce small deviations from GR for scales $k\lesssim k_{nr}$ and a considerable enhancement for $k\gtrsim k_{nr}$. This constraint already introduces a preferred value for the braiding scale $k_B$, which has to be similar to $k_{nr}$. In general one could tune the ratio $\alpha_K/\alpha^2_B$ in order to place the transition between the two gravity regimes at the desired scale, but in this case the amplitude of $\alpha_B$ is already fixed by the neutrino mass, since it is the only parameter that can enhance the power spectrum at large \textit{k}. Therefore the only possible way to match the two scales is to change the value of $\alpha_K$. Note however that, \textit{at linear level}, different values of $\alpha_K$ do not change matter power spectrum features but just the range of modes where these features appear.
For bigger kineticity values the braiding scale is translated to larger $k$ as well as the desired MG step-like feature, however as soon as $k_B>k_{nr}$ there is a range of scales where we start to observe the neutrino-induced suppression but not the MG-induced enhancement. As a result the fit favours a larger primordial fluctuations amplitude (see right column of Table \ref{tab:parameters}).

Given this particular signature due to the kineticity, we have done additional runs where also $\alpha_K$ was allowed to vary (assuming a uniform prior) while keeping fixed the modulation function parameters to $(z_{th},\Delta z)=(100,20)$ and the neutrino model to \texttt{M500}, finding for the first time constraints on this parameter: $\alpha_K$ is peaked around $\alpha_K^{peak}=6.5$ and is bounded to be in the interval $[1.9,20.4]$ at the $95\%$ CL.

Naively this is unexpected, since in the quasi-static (QS) limit --i.e. for modes such that $k/k_{c_s}\gg 1$, where $k_{c_s}=\frac{aH}{c_s}$ is the dark-energy sound horizon \cite{sawicki:qsapprox}-- the kineticity $\alpha_K$ disappears from the perturbed equations. However, an Euclid-like survey will probe considerably large scales  where the QS approximation may not hold. In fact,
\begin{equation}
\frac{k}{k_{c_s}}\geq\frac{k_{survey,min}}{k_{c_s}(z=2)}\sim 0.3,
\end{equation}
indicating that the QS limit condition is not fulfilled for every mode and redshift.

\section{Removing degeneracies}

In order to explore the degeneracy between modified gravity parameters and neutrino mass, we performed additional MCMC runs with different datasets and two different fiducial models (see Table \ref{tab:MCMC}). We used the \texttt{GR+M0} fiducial described in the previous section and a new  fiducial model: \texttt{GR+M150}; this has the same cosmological parameters of \texttt{GR+M0} but it includes the \texttt{M150} neutrino mass model. In both cases we fixed the values of $(c_K,c_T,z_{th},\Delta z)$ to $(10,0,100,50)$, choosing this value for the tensor speed excess because its impact on the growth of structures is negligible. In each chain the varying parameters were the six standard cosmological parameters, the coefficients of the braiding and the Planck mass run-rate ($c_B$ and $c_M$) and the neutrino mass $\Sigma m_\nu$, under the simplifying assumptions of normal mass ordering. For each varying parameter we assumed uniform priors.

\begin{figure}
\centerline{
\includegraphics[scale=0.8]{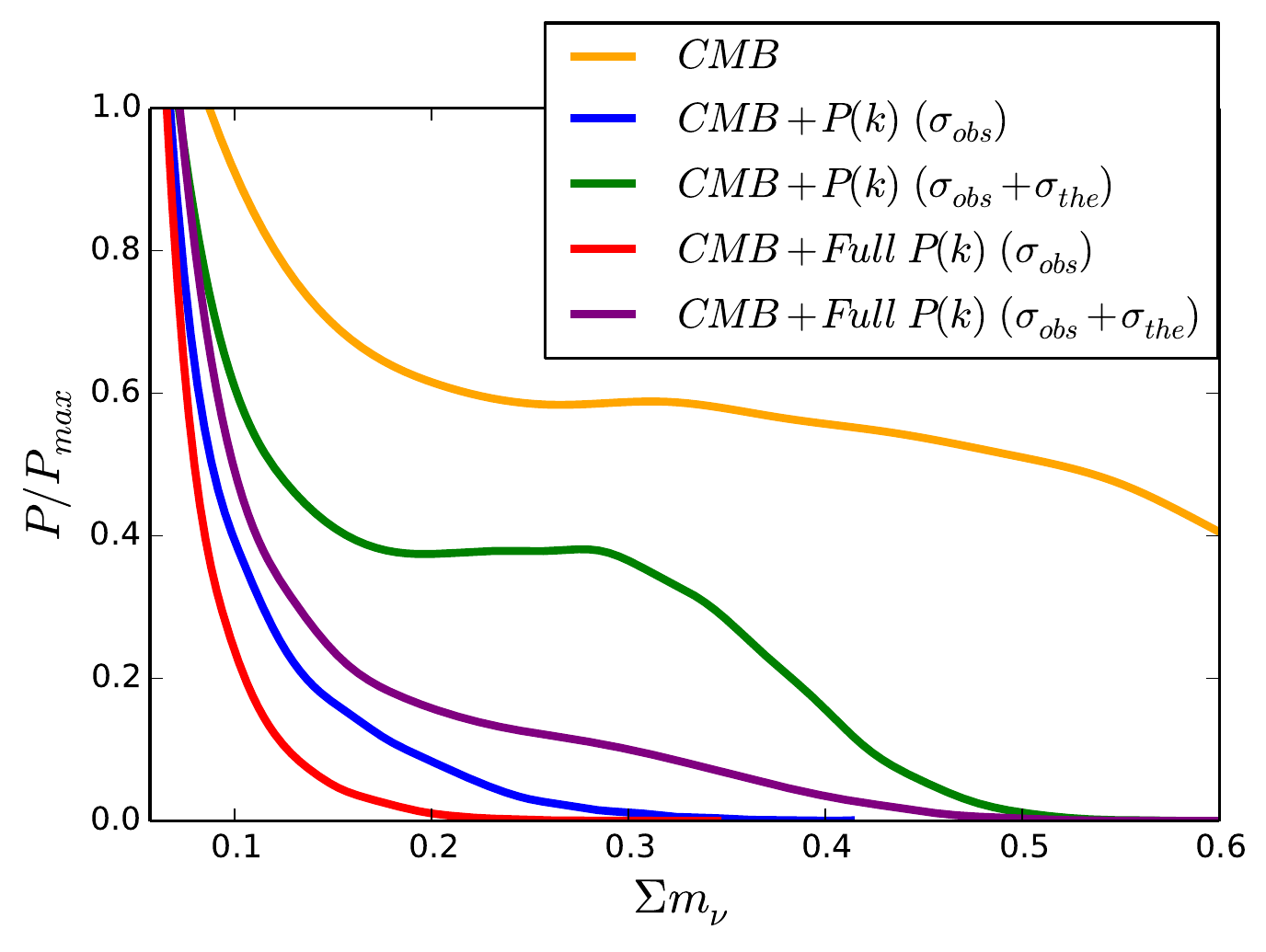}} % 80mm
\caption{Neutrino mass posterior (relative to its maximum) of modified gravity models for several datasets. The fiducial is given by the \texttt{M0} model. The posterior wide tail of the \textit{CMB} dataset is considerably damped when power spectrum data is added. Significantly, heavy neutrinos, considered in the previous section, are highly disfavoured once we include information coming from the expansion history.
\label{figure:step2_mnu_posterior}}
\end{figure}

In Figure \ref{figure:step2_mnu_posterior} we show the posterior distributions for the \texttt{M0} fiducial and different datasets. The broad tail in the \textit{CMB} posterior indicates that this dataset alone is not able to tightly constrain neutrino mass, allowing neutrinos to be quite massive ($\sum m_\nu \lesssim 0.761\ eV$ at $95\%$ CL). We also considered power spectra datasets, where we implemented different combinations of observational error $\sigma_{obs}$, theoretical error $\sigma_{the}$ and effects coming from changes in the expansion history. Neutrino mass constraints come mainly from the small scales, where few percent observational errors lead to tight constraints. On the other hand, non-linear effects and modelling systematics both in neutrinos and modified gravity physics appear precisely at these scales, so even a theoretical error rough estimate, as adopted here, can show how our constraints weaken. Comparing curves with and without the theoretical error we can see that, if we don't take into account effects coming from different expansion histories (\textit{CMB+P(k) ($\sigma_{obs}+\sigma_{the}$)} dataset), inaccurate modelling at small scales could really modify the high mass tail of the distribution ($\sum m_\nu \lesssim 0.400\ eV$ at $95\%$ CL). The expansion history information enclosed in \textit{Full P(k)}  is able to play a significant role in constraining neutrino mass (compare \textit{CMB+Full P(k)} to \textit{CMB+P(k)})\textcolor{blue}, but these constraints could weaken by changing the Hubble expansion rate $H(z)$ or the angular diameter distance $D_A(z)$ through a suitable choice of equation of state parameter for dark energy, away from the cosmological constant value, $w_{\phi}(z)\neq -1$.

\begin{figure}
\centerline{
\includegraphics[scale=0.8]{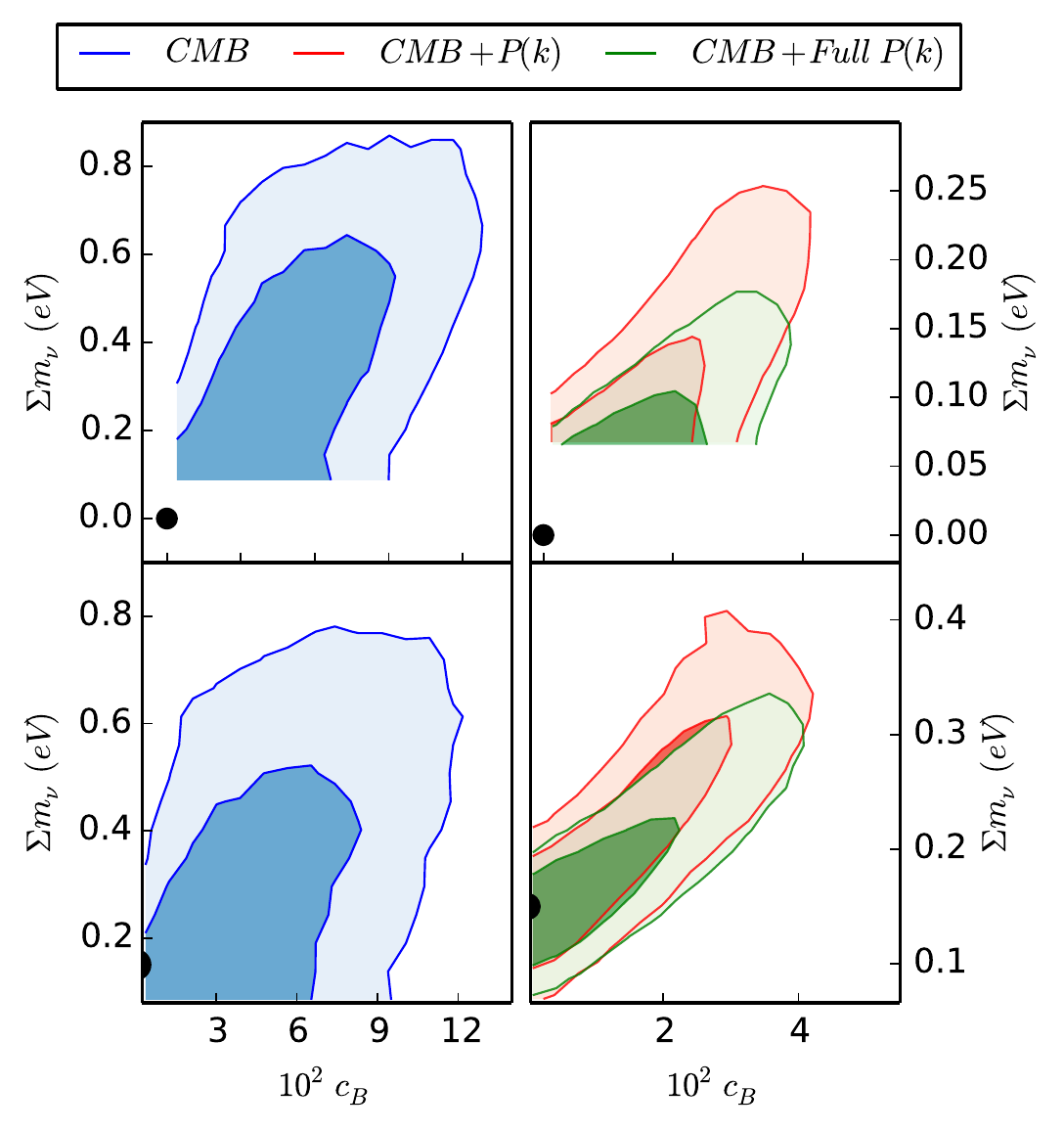}} % 80mm
\caption{Correlation between $c_B$ and $\Sigma m_\nu$ assuming a fiducial model \texttt{GR+M0} (upper panels) or \texttt{GR+M150} (lower panels). Fiducials are represented as black dots. Contours are shown for the $68\%$ and $95\%$ CL when considering \textit{CMB} (left panels), \textit{CMB+P(k)} and \textit{CMB+Full P(k)} (right panels) datasets. In the latter case we have plotted the case with only observational errors, which is the most constraining case. Cuts in the correlations appear as a result of our assumptions: in our model $c_B$ cannot be negative without triggering instabilities and $\Sigma m_\nu$ cannot be less than $\simeq 0.06$ eV since we imposed that neutrinos obey the normal ordering mass scheme.}
\label{figure:step2_mg_degeneracy}
\end{figure}

In the case of the \texttt{M0} fiducial model, considering the most constraining dataset, MG can hide neutrino masses up to 160 meV at $95\%$ CL. On the contrary, a large neutrino mass compatible with present bounds (\texttt{M150} fiducial model) can not be hidden by MG as shown in Figure \ref{figure:step2_mg_degeneracy}, where the analysis with the largest dataset finds a minimum mass of 100 meV at $95\%$ CL.

Our results show that  the neutrino mass is degenerate with the $c_B$ parameter, as illustrated in Figure \ref{figure:step2_mg_degeneracy} for the two fiducial models \texttt{GR+M0} and \texttt{GR+M150}. $c_B$ is the only parameter (at the linear level) in our models that can hide the suppression induced in the power spectrum by massive neutrinos. We can draw this conclusion also by looking at the \textit{CMB} MCMC result, since we included in the dataset also the power spectrum of the weak lensing CMB deflection angle, which in turns depends on the gravitational potentials along the line of sight. Adding power spectrum observations also shows the $c_B-\sum m_\nu$ correlation, with a neutrino mass range partially limited.

\section{Conclusions}
In this work, we have explored the nature of the degeneracy between neutrino masses and modified gravity. We focussed on widely applicable results by using the Horndeski generalized scalar-tensor theories of gravity.  These generic models contain, in the linear regime, few redshift dependent functions that here are tuned to maximize the impact on the neutrino mass determination. We have studied the parameter space of the number of modified gravity parameters (6) in the model for several neutrino mass schemes and considered  mock state-of-the-art CMB data sets and forthcoming  galaxy  redshift surveys.  Among other minor contributions, one parameter describing one of the four Hordenski functions is the dominant source of degeneracy with neutrino masses. 
Not unexpectedly this is the braiding, which physically arises from  a mixing between the kinetic terms of the metric and the scalar and it modifies the growth of perturbations  boosting small scale power.

We have shown the cancellation of the impact of the neutrino mass with a modified gravity model in the power spectrum in the linear regime. This cancellation is very efficient at a particular redshift; combinations of several redshifts may lift the degeneracy and we have explored the potential of future datasets in doing so. We find that future data such as those provided by an Euclid-like survey  would limit, but not fully cancel, the degeneracy of neutrino mass with one of the Hordenski parameters, $c_B$. 
Here we have only considered galaxy redshift surveys as large-scale structure probes. Of course the weak gravitational lensing signal could further help, but consideration of this probe is left to future work.

\begin{acknowledgments}
NB is supported by the Spanish MINECO under grant BES-2015-073372. BH is partially supported by a Beatriu de Pinos grant and the Chinese National Youth Thousand Talents Program. We acknowledge support by Spanish Mineco grants AYA2014-58747-P, FPA2014-57816-P, GVA grant PROMETEOII/2014/050, MDM-2014-0369 of ICCUB (Unidad de Excelencia Maria de Maeztu) and  EU projects H2020-MSCA-ITN-2015//674896-ELUSIVES  and H2020-MSCA-RISE-2015.
\end{acknowledgments}

%\bibliography{biblio}

\begin{thebibliography}{10}

\bibitem{Ade:2015xua}
{\bfseries Planck} Collaboration, P.~A.~R. Ade {\em et~al.}, ``{Planck 2015
  results. XIII. Cosmological parameters},''
  \href{http://dx.doi.org/10.1051/0004-6361/201525830}{{\em Astron. Astrophys.}
  {\bfseries 594} (2016) A13},
\href{http://arxiv.org/abs/1502.01589}{{\ttfamily arXiv:1502.01589
  [astro-ph.CO]}}.
%%CITATION = ARXIV:1502.01589;%%.

\bibitem{Ade:2015rim}
{\bfseries Planck} Collaboration, P.~A.~R. Ade {\em et~al.}, ``{Planck 2015
  results. XIV. Dark energy and modified gravity},''
  \href{http://dx.doi.org/10.1051/0004-6361/201525814}{{\em Astron. Astrophys.}
  {\bfseries 594} (2016) A14},
\href{http://arxiv.org/abs/1502.01590}{{\ttfamily arXiv:1502.01590
  [astro-ph.CO]}}.
%%CITATION = ARXIV:1502.01590;%%.

\bibitem{Clifton:2011jh}
T.~Clifton, P.~G. Ferreira, A.~Padilla, and C.~Skordis, ``{Modified Gravity and
  Cosmology},'' \href{http://dx.doi.org/10.1016/j.physrep.2012.01.001}{{\em
  Phys. Rept.} {\bfseries 513} (2012) 1--189},
\href{http://arxiv.org/abs/1106.2476}{{\ttfamily arXiv:1106.2476
  [astro-ph.CO]}}.
%%CITATION = ARXIV:1106.2476;%%.

\bibitem{Kristiansen2010609}
J.~Kristiansen, G.~L. Vacca, L.~Colombo, R.~Mainini, and S.~Bonometto,
  ``Coupling between cold dark matter and dark energy from neutrino mass
  experiments,''
  \href{http://dx.doi.org/http://dx.doi.org/10.1016/j.newast.2010.02.003}{{\em
  New Astronomy} {\bfseries 15} no.~7, (2010) 609 -- 613},
  \href{http://arxiv.org/abs/arXiv:0902.2737}{{\ttfamily arXiv:0902.2737}}.
  \url{http://www.sciencedirect.com/science/article/pii/S1384107610000138}.

\bibitem{lavacca:cdmdeneutrinos}
G.~L. Vacca, J.~Kristiansen, L.~Colombo, R.~Mainini, and S.~Bonometto, ``Do
  WMAP data favor neutrino mass and a coupling between Cold Dark Matter and
  Dark Energy?,'' {\em Journal of Cosmology and Astroparticle Physics}
  {\bfseries 2009} no.~04, (2009) 007,
  \href{http://arxiv.org/abs/arXiv:0902.2711}{{\ttfamily arXiv:0902.2711}}.
  \url{http://stacks.iop.org/1475-7516/2009/i=04/a=007}.

\bibitem{lavacca:ddeneutrinos}
G.~L. Vacca and J.~Kristiansen, ``Dynamical Dark Energy model parameters with
  or without massive neutrinos,'' {\em Journal of Cosmology and Astroparticle
  Physics} {\bfseries 2009} no.~07, (2009) 036,
  \href{http://arxiv.org/abs/arXiv:0906.4501}{{\ttfamily arXiv:0906.4501}}.
  \url{http://stacks.iop.org/1475-7516/2009/i=07/a=036}.

\bibitem{Baldi:2013iza}
M.~Baldi, F.~Villaescusa-Navarro, M.~Viel, E.~Puchwein, V.~Springel, and
  L.~Moscardini, ``{Cosmic degeneracies ? I. Joint N-body simulations of
  modified gravity and massive neutrinos},''
  \href{http://dx.doi.org/10.1093/mnras/stu259}{{\em Mon. Not. Roy. Astron.
  Soc.} {\bfseries 440} no.~1, (2014) 75--88},
\href{http://arxiv.org/abs/1311.2588}{{\ttfamily arXiv:1311.2588
  [astro-ph.CO]}}.
%%CITATION = ARXIV:1311.2588;%%.

\bibitem{Barreira:2014ija}
A.~Barreira, B.~Li, C.~Baugh, and S.~Pascoli, ``{Modified gravity with massive
  neutrinos as a testable alternative cosmological model},''
  \href{http://dx.doi.org/10.1103/PhysRevD.90.023528}{{\em Phys. Rev.}
  {\bfseries D90} no.~2, (2014) 023528},
\href{http://arxiv.org/abs/1404.1365}{{\ttfamily arXiv:1404.1365
  [astro-ph.CO]}}.
%%CITATION = ARXIV:1404.1365;%%.

\bibitem{Shim:2014uta}
J.~Shim, J.~Lee, and M.~Baldi, ``{Breaking the Cosmic Degeneracy between
  Modified Gravity and Massive Neutrinos with the Cosmic Web},''
\href{http://arxiv.org/abs/1404.3639}{{\ttfamily arXiv:1404.3639
  [astro-ph.CO]}}.
%%CITATION = ARXIV:1404.3639;%%.

\bibitem{Horndeski:1974wa}
G.~W. Horndeski, ``{Second-order scalar-tensor field equations in a
  four-dimensional space},''
\href{http://dx.doi.org/10.1007/BF01807638}{{\em Int. J. Theor. Phys.}
  {\bfseries 10} (1974) 363--384}.
%%CITATION = IJTPB,10,363;%%.

\bibitem{Deffayet:2009mn}
C.~Deffayet, S.~Deser, and G.~Esposito-Farese, ``{Generalized Galileons: All
  scalar models whose curved background extensions maintain second-order field
  equations and stress-tensors},''
  \href{http://dx.doi.org/10.1103/PhysRevD.80.064015}{{\em Phys. Rev.}
  {\bfseries D80} (2009) 064015},
\href{http://arxiv.org/abs/0906.1967}{{\ttfamily arXiv:0906.1967 [gr-qc]}}.
%%CITATION = ARXIV:0906.1967;%%.

\bibitem{Kobayashi:2011nu}
T.~Kobayashi, M.~Yamaguchi, and J.~Yokoyama, ``{Generalized G-inflation:
  Inflation with the most general second-order field equations},''
  \href{http://dx.doi.org/10.1143/PTP.126.511}{{\em Prog. Theor. Phys.}
  {\bfseries 126} (2011) 511--529},
\href{http://arxiv.org/abs/1105.5723}{{\ttfamily arXiv:1105.5723 [hep-th]}}.
%%CITATION = ARXIV:1105.5723;%%.

\bibitem{Gubitosi:2012hu}
G.~Gubitosi, F.~Piazza, and F.~Vernizzi, ``{The Effective Field Theory of Dark
  Energy},'' \href{http://dx.doi.org/10.1088/1475-7516/2013/02/032}{{\em JCAP}
  {\bfseries 1302} (2013) 032},
  \href{http://arxiv.org/abs/1210.0201}{{\ttfamily arXiv:1210.0201 [hep-th]}}.
[JCAP1302,032(2013)].
%%CITATION = ARXIV:1210.0201;%%.

\bibitem{Bloomfield:2012ff}
J.~K. Bloomfield, E.~E. Flanagan, M.~Park, and S.~Watson, ``{Dark energy or
  modified gravity? An effective field theory approach},''
  \href{http://dx.doi.org/10.1088/1475-7516/2013/08/010}{{\em JCAP} {\bfseries
  1308} (2013) 010},
\href{http://arxiv.org/abs/1211.7054}{{\ttfamily arXiv:1211.7054
  [astro-ph.CO]}}.
%%CITATION = ARXIV:1211.7054;%%.

\bibitem{Gleyzes:2013ooa}
J.~Gleyzes, D.~Langlois, F.~Piazza, and F.~Vernizzi, ``{Essential Building
  Blocks of Dark Energy},''
  \href{http://dx.doi.org/10.1088/1475-7516/2013/08/025}{{\em JCAP} {\bfseries
  1308} (2013) 025},
\href{http://arxiv.org/abs/1304.4840}{{\ttfamily arXiv:1304.4840 [hep-th]}}.
%%CITATION = ARXIV:1304.4840;%%.

\bibitem{Hu:2013twa}
B.~Hu, M.~Raveri, N.~Frusciante, and A.~Silvestri, ``{Effective Field Theory of
  Cosmic Acceleration: an implementation in CAMB},''
  \href{http://dx.doi.org/10.1103/PhysRevD.89.103530}{{\em Phys. Rev.}
  {\bfseries D89} no.~10, (2014) 103530},
\href{http://arxiv.org/abs/1312.5742}{{\ttfamily arXiv:1312.5742
  [astro-ph.CO]}}.
%%CITATION = ARXIV:1312.5742;%%.

\bibitem{bellini:horndeski}
E.~Bellini and I.~Sawicki, ``Maximal freedom at minimum cost: linear
  large-scale structure in general modifications of gravity,'' {\em Journal of
  Cosmology and Astroparticle Physics} {\bfseries 2014} no.~07, (2014) 050,
  \href{http://arxiv.org/abs/arXiv:1404.3713v3}{{\ttfamily arXiv:1404.3713v3}}.
  \url{http://stacks.iop.org/1475-7516/2014/i=07/a=050}.

\bibitem{Linder:2015rcz}
E.~V. Linder, G.~Sengor, and S.~Watson, ``{Is the Effective Field Theory of
  Dark Energy Effective?},''
  \href{http://dx.doi.org/10.1088/1475-7516/2016/05/053}{{\em JCAP} {\bfseries
  1605} no.~05, (2016) 053},
\href{http://arxiv.org/abs/1512.06180}{{\ttfamily arXiv:1512.06180
  [astro-ph.CO]}}.
%%CITATION = ARXIV:1512.06180;%%.

\bibitem{bellini:constraints}
E.~Bellini, A.~J. Cuesta, R.~Jimenez, and L.~Verde, ``Constraints on deviations
  from $\Lambda$CDM within Horndeski gravity,'' {\em Journal of Cosmology and
  Astroparticle Physics} {\bfseries 2016} no.~02, (2016) 053,
  \href{http://arxiv.org/abs/arXiv:1509.07816v3}{{\ttfamily
  arXiv:1509.07816v3}}. \url{http://stacks.iop.org/1475-7516/2016/i=02/a=053}.

\bibitem{Alonso:2016suf}
D.~Alonso, E.~Bellini, P.~G. Ferreira, and M.~Zumalacarregui, ``{The
  Observational Future of Cosmological Scalar-Tensor Theories},''
\href{http://arxiv.org/abs/1610.09290}{{\ttfamily arXiv:1610.09290
  [astro-ph.CO]}}.
%%CITATION = ARXIV:1610.09290;%%.

\bibitem{perenon:horndeski}
L.~Perenon, C.~Marinoni, and F.~Piazza, ``Diagnostic of Horndeski Theories,''
  \href{http://arxiv.org/abs/arXiv:1609.09197v2}{{\ttfamily
  arXiv:1609.09197v2}}.

\bibitem{Planck:fiducial}
{\bfseries Planck} Collaboration, ``Planck intermediate results. XLVI.
  Reduction of large-scale systematic effects in HFI polarization maps and
  estimation of the reionization optical depth,''
  \href{http://arxiv.org/abs/arXiv:1605.02985}{{\ttfamily arXiv:1605.02985}}.

\bibitem{Planck:2006aa}
{\bfseries Planck} Collaboration, ``The Scientific Programme of Planck,''
  \href{http://arxiv.org/abs/arXiv:astro-ph/0604069}{{\ttfamily
  arXiv:astro-ph/0604069}}.

\bibitem{Lewis:1999bs}
A.~Lewis, A.~Challinor, and A.~Lasenby, ``Efficient Computation of Cosmic
  Microwave Background Anisotropies in Closed Friedmann-Robertson-Walker
  Models,'' {\em The Astrophysical Journal} {\bfseries 538} no.~2, (2000) 473,
  \href{http://arxiv.org/abs/arXiv:astro-ph/9911177}{{\ttfamily
  arXiv:astro-ph/9911177}}.
  \url{http://stacks.iop.org/0004-637X/538/i=2/a=473}.

\bibitem{Perotto:2006rj}
L.~Perotto, J.~Lesgourgues, S.~Hannestad, H.~Tu, and Y.~Y.~Y. Wong, ``{Probing
  cosmological parameters with the CMB: Forecasts from full Monte Carlo
  simulations},'' \href{http://dx.doi.org/10.1088/1475-7516/2006/10/013}{{\em
  JCAP} {\bfseries 0610} (2006) 013},
\href{http://arxiv.org/abs/astro-ph/0606227}{{\ttfamily arXiv:astro-ph/0606227
  [astro-ph]}}.
%%CITATION = ASTRO-PH/0606227;%%.

\bibitem{Hu:2001kj}
T.~Okamoto and W.~Hu, ``Cosmic microwave background lensing reconstruction on
  the full sky,'' \href{http://dx.doi.org/10.1103/PhysRevD.67.083002}{{\em
  Phys. Rev. D} {\bfseries 67} (Apr, 2003) 083002},
  \href{http://arxiv.org/abs/arXiv:astro-ph/0301031}{{\ttfamily
  arXiv:astro-ph/0301031}}.

\bibitem{Lewis:2005tp}
A.~Lewis, ``{Lensed CMB simulation and parameter estimation},''
  \href{http://dx.doi.org/10.1103/PhysRevD.71.083008}{{\em Phys. Rev.}
  {\bfseries D71} (2005) 083008},
\href{http://arxiv.org/abs/astro-ph/0502469}{{\ttfamily arXiv:astro-ph/0502469
  [astro-ph]}}.
%%CITATION = ASTRO-PH/0502469;%%.

\bibitem{euclid:databook}
L.~Amendola and E.~T.~W. Group, ``Cosmology and Fundamental Physics with the
  Euclid Satellite,'' \href{http://dx.doi.org/10.1007/lrr-2013-6}{{\em Living
  Reviews in Relativity} {\bfseries 16} no.~6, (2013) },
  \href{http://arxiv.org/abs/arXiv:1206.1225v2}{{\ttfamily arXiv:1206.1225v2}}.

\bibitem{audren:forecast}
B.~Audren, J.~Lesgourgues, S.~Bird, M.~G. Haehnelt, and M.~Viel, ``Neutrino
  masses and cosmological parameters from a Euclid-like survey: Markov Chain
  Monte Carlo forecasts including theoretical errors,'' {\em Journal of
  Cosmology and Astroparticle Physics} {\bfseries 2013} no.~01, (2013) 026,
  \href{http://arxiv.org/abs/arXiv:1210.2194v1}{{\ttfamily arXiv:1210.2194v1}}.
  \url{http://stacks.iop.org/1475-7516/2013/i=01/a=026}.

\bibitem{zumalacarregui:hiclass}
M.~Zumalacarregui, E.~Bellini, I.~Sawicki, and J.~Lesgourgues,
  ``\texttt{hi\_class}: Horndeski in the Cosmic Linear Anisotropy Solving
  System,'' \href{http://arxiv.org/abs/arXiv:1605.06102}{{\ttfamily
  arXiv:1605.06102}}.

\bibitem{lesgourgues:class}
J.~Lesgourgues, ``The Cosmic Linear Anisotropy Solving System (CLASS) I:
  Overview,'' \href{http://arxiv.org/abs/arXiv:1104.2932}{{\ttfamily
  arXiv:1104.2932}}.

\bibitem{blas:class}
D.~Blas, J.~Lesgourgues, and T.~Tram, ``The Cosmic Linear Anisotropy Solving
  System (CLASS). Part II: Approximation schemes,'' {\em Journal of Cosmology
  and Astroparticle Physics} {\bfseries 2011} no.~07, (2011) 034,
  \href{http://arxiv.org/abs/arXiv:1104.2933}{{\ttfamily arXiv:1104.2933}}.
  \url{http://stacks.iop.org/1475-7516/2011/i=07/a=034}.

\bibitem{lesgourgues:class_ncdm}
J.~Lesgourgues and T.~Tram, ``The Cosmic Linear Anisotropy Solving System
  (CLASS) IV: efficient implementation of non-cold relics,'' {\em Journal of
  Cosmology and Astroparticle Physics} {\bfseries 2011} no.~09, (2011) 032,
  \href{http://arxiv.org/abs/arXiv:1104.2935}{{\ttfamily arXiv:1104.2935}}.
  \url{http://stacks.iop.org/1475-7516/2011/i=09/a=032}.

\bibitem{audren:montepython}
B.~Audren, J.~Lesgourgues, K.~Benabed, and S.~Prunet, ``Conservative
  constraints on early cosmology with MONTE PYTHON,'' {\em Journal of Cosmology
  and Astroparticle Physics} {\bfseries 2013} no.~02, (2013) 001,
  \href{http://arxiv.org/abs/arXiv:1210.7183}{{\ttfamily arXiv:1210.7183}}.
  \url{http://stacks.iop.org/1475-7516/2013/i=02/a=001}.

\bibitem{sawicki:qsapprox}
I.~Sawicki and E.~Bellini, ``Limits of quasistatic approximation in
  modified-gravity cosmologies,''
  \href{http://dx.doi.org/10.1103/PhysRevD.92.084061}{{\em Phys. Rev. D}
  {\bfseries 92} (Oct, 2015) 084061},
  \href{http://arxiv.org/abs/arXiv:1503.06831v2}{{\ttfamily
  arXiv:1503.06831v2}}.

\end{thebibliography}
%\bibliographystyle{utcaps}

\providecommand{\href}[2]{#2}\begingroup\raggedright\endgroup

\end{document}